\documentclass[10pt,conference]{IEEEtran}
\IEEEoverridecommandlockouts
\usepackage{cite}
\usepackage{amsmath,amssymb,amsfonts}
\usepackage{algorithmic}
\usepackage{graphicx}
\usepackage{balance}
\usepackage{textcomp}
\usepackage{xcolor}
\usepackage{tabularx}
\usepackage{multirow}
\usepackage{booktabs}
\usepackage{hyperref}
\usepackage[multiple]{footmisc}
\usepackage{changepage} % for adjustwidth environment
\usepackage{caption} % for formatting caption
\def\BibTeX{{\rm B\kern-.05em{\sc i\kern-.025em b}\kern-.08em
    T\kern-.1667em\lower.7ex\hbox{E}\kern-.125emX}}
\begin{document}

\title{SoK: Trusted Execution in SoC-FPGAs}

\author{\IEEEauthorblockN{
Garrett Perkins\IEEEauthorrefmark{1},
Benjamin Macht\IEEEauthorrefmark{1},
Lucas Ritzdorf\IEEEauthorrefmark{1},
Tristan Running Crane\IEEEauthorrefmark{1},\\
Brock LaMeres\IEEEauthorrefmark{1},
Clemente Izurieta\IEEEauthorrefmark{1}\IEEEauthorrefmark{2}\IEEEauthorrefmark{3}, 
Ann Marie Reinhold\IEEEauthorrefmark{1}\IEEEauthorrefmark{3}\\[2.0ex]
}

\IEEEauthorblockA{\IEEEauthorrefmark{1}Montana State University, Bozeman, MT, USA}  
\IEEEauthorblockA{\IEEEauthorrefmark{2}Idaho National Laboratory, Idaho Falls, ID, USA \hspace{1em}
\IEEEauthorrefmark{3}Pacific Northwest National Laboratory, Richland, WA, USA}
}

\maketitle

\begin{abstract}
Trusted Execution Environments (TEEs) have emerged at the forefront of edge computing to combat the lack of trust between system components. Field Programmable Gate Arrays (FPGAs) are commonly used as edge computers but were not created with security as a primary consideration. Thus, FPGA-based edge computers are increasingly the target of cyberattacks. We analyze the existing literature to systematize the applications and features of FPGA-based TEEs. We identified 27 primary studies related to different types of System-on-Chip FPGA-based TEEs. Across a wide range of applications and features, the availability of extensible solutions is limited. Most solutions focus on specific features and applications, whereas few solutions focus on feature-rich, comprehensive TEEs that can be utilized across computer systems. Whether TEEs are specific or extensible, the paucity of published studies provides evidence of research gaps. This SoK delineates these gaps revealing opportunities for researchers and developers.  
\end{abstract}


\section{Introduction}
In the rapidly evolving landscape of the Internet of Things (IoT) and edge computing, the demand for secure environments (SEs) has grown markedly. With the increasing interconnectivity of devices, traditional computer systems are no longer able to rely on mutual trust among components, as a compromise in one area can lead to vulnerabilities in others \cite{PSTcloud}. This heightened risk has underscored the need for SEs that can adapt to the challenges posed by the evolving domain of secure computing \cite{sec_edge_chall, sec_edge_survey, secPriv_issues}.

Most major CPU vendors have introduced their own chip-specific Trusted Execution Environments (TEE) solutions. For example, \textit{ARM TrustZone}, \textit{Intel SGX}, and \textit{AMD SEV}, each provide secure computing for their respective hardware. However, these chip-specific TEEs constrain developers to a singular platform creating a unique security challenge \cite{kaplan2017protecting, keystone, arm_developer_202}.

New solutions are emerging to address this challenge by providing more modular and flexible secure environments.  Consequently, significant R\&D efforts are being applied to TEEs and hardware-based solutions. Among these hardware solutions are FPGA and ASICs. This paper focuses on FPGAs, which are application-agnostic, as opposed to ASICs, which are ``application-specific" by definition. FPGAs are inherently modular because of their field programmability and used across several applications, such as radar, Unmanned Aerial Vehicles (UAVs), Industrial Control Systems (ICS), data centers, neural networks, and space avionics \cite{Arm_FPGA, Schneider_Smalley_2024, Intel_FPGA_ASIC}. These applications require that FPGAs be secure.

We explore how FPGA-based TEEs are currently being used to provide secure computing environments and the specific features that make them suitable for applications in IoT and other computing domains. By highlighting gaps in existing research and solutions that improve FPGA security, our study addresses the following research questions: \textbf{RQ1:} What are the applications of FPGA-based TEEs and which features do FPGA-based TEEs employ according to the literature? \textbf{RQ2:} What gaps exist in the field of FPGA-based TEEs according to the literature?

\section{Methods}
We searched two databases, ACM Digital Library and IEEE Xplore, identifying 109 peer-reviewed papers using the search strings and filters shown in Table \ref{fig:papers_over_time}. After applying inclusion criteria (Table \ref{table:criteria}), 27 papers remained for full evaluation.

We applied inclusion criteria focused on the convergence of TEEs, FPGAs, and cybersecurity. First, we included only papers that primarily addressed security concerns, excluding those not focused on security. Second, we considered only studies that demonstrated practical implementations or empirical evaluations, thus excluding theoretical papers and literature reviews. Furthermore, our review was limited to papers discussing System-on-Chip (SoC)-based FPGA environments, excluding those involving non-SoC processors to maintain technological specificity. Last, we prioritized open-source systems, excluding studies reliant on proprietary platforms. This prioritization ensured the studies were universally accessible and modifiable. This meticulous selection process was critical to accurately mapping the landscape of FPGA-based TEEs, identifying their applications, and detailing the specific features they employ, directly addressing our research question.

Of the 109 papers, 75 were from ACM Digital Library, and 34 were from IEEE Xplore. After applying the inclusion criteria listed in Table \ref{table:criteria}, 31 papers remained: 17 from IEEE Xplore and 14 from ACM Digital Library. Despite meeting the inclusion criteria, four papers were removed from the pool of 31 due to lack of relevance, leaving 27 papers in the study. A stacked bar plot was made based on the number of papers published each year (Figure \ref{fig:papers_over_time}). 

After selecting 27 papers, each was read to categorize the features and applications of these custom TEEs. Notable features and applications were separately categorized by paper in Table \ref{table:applications&features}. This table does not include the papers \cite{keystone, HECTOR-V}, and \cite{keystone-sensor}, as \cite{keystone} and \cite{HECTOR-V} are categorized as extensible TEEs and \cite{keystone-sensor} is an implementation of \cite{keystone}.

We built a heatmap to identify which features are most commonly associated with each application and highlight areas of researcher attention (Figure \ref{fig:heatmap}). This aids in visualizing the distribution of features across various applications of FPGA-based TEEs and facilitates clear and immediate understanding of the landscape of FPGA-based TEEs.

\begin{table}[t]
\captionsetup{format=plain, singlelinecheck=false, justification=raggedright}
\caption{Database search details, including search strings, filters, results.}
\centering
\renewcommand{\arraystretch}{1.2} % Adjusts row height for better readability
\begin{tabularx}{\columnwidth}{|p{1cm}|p{3cm}|p{2cm}|p{1.1cm}|}  
 \hline
 \textbf{Database} & \textbf{Search String} & \textbf{Filters} & \textbf{Results} \\
 \hline
 ACM Digital Library & [``trusted execution environment"] AND [fpga] & Past 5 years, Research Articles Only & 75 \\
 \hline
 IEEE Xplore & (``All Metadata":``trusted execution environment") AND (``All Metadata":fpga) & 2019-2024, Journals/Conferences & 34 \\
 \hline
\end{tabularx}
\label{table:strings}
\end{table}

\begin{table}[t]
\centering
\caption{Inclusion criteria and number of papers excluded for each criterion. 
Total count of excluded papers exceeds the 109 papers obtained from the initial search strings because some papers were excluded for not meeting multiple criteria.}
\begin{tabular}[t]{l c}
\toprule
\textbf{Criteria} & \textbf{Count of papers excluded} \\
\midrule
Security Focused      & 4  \\
Applied Research     & 16 \\
Open-source Platform & 28 \\
System on Chip Based & 63 \\
\bottomrule
\end{tabular}
\label{table:criteria}
\end{table}

\section{Results \& Discussion}

 The increasing rate of publications around TEEs and FPGAs indicates that these are both growing areas of research in the cybersecurity community (Figure \ref{fig:papers_over_time})\cite{kitchenham2011using}. Despite recent growth in publications, research on FPGA-based TEEs remains limited, with only 27 relevant studies identified.

The majority of the 27 papers focus on applications and features. More specifically, 24 papers address application-specific (15 papers) (Subsection \ref{app}) and feature-specific (9 papers)(Subsection \ref{features}) TEEs. Only two papers present a holistic approach to TEEs (Subsection \ref{extensible}). One paper presents a use case of a holistic approach. The application-based papers focus on the topics of accelerators, cloud computing, and attack mitigation (Table \ref{table:applications&features}); the rest of the 24 papers are feature-driven. Root of Trust (RoT) and various memory security features are common, while features such as password recovery and upgraded page table walks are less common. Each paper presents a unique combination of applications and features (Figure \ref{fig:heatmap}).

\subsection{Application Specific}
\label{app}
Across the 27 selected papers, 15 constructed TEEs that served niche purposes. However, note that some of these "niche papers" developed TEEs that are multi-applicational (i.e., acceleration in cloud computing) but not fully extensible. 

\subsubsection{Hardware Acceleration}
\label{accel}
Seven papers discuss custom TEEs and their features as they are applied to accelerators and acceleration. TEEs emphasizing hardware acceleration primarily feature memory security, enclaves, RoT, and attestation (Figure \ref{fig:heatmap}). \textit{ShEF} implements a unique Shield module for secure data access \cite{ShEF}. Meanwhile, \textit{TACC} separates memory management for in-package (internal) and off-package (external) memory \cite{TACC}. \textit{AccGuard} separates and isolates memory regions for use in multi-tenant cloud environments, whereas  \textit{AccShield} supports unified virtual memory across multiple accelerators, allowing them to securely share memory resources \cite{AccShield}. Paper \cite{Software-defined} used a Software-Defined Interconnect block, a hardware block that dynamically controls and sets specific boundaries for memory regions.  While secure memory is the most widespread hardware acceleration feature, other features are discussed in the literature.

Papers \cite{ShEF}, \cite{TACC}, and \cite{AccGuard} differ on cloud-specific use cases, but all take an enclave-based approach to TEEs. Papers \cite{ShEF}, \cite{AccGuard}, and \cite{AccShield} required attestation with a root of trust for verification purposes. Other features were less prevalent across papers focused on hardware acceleration (e.g., Secure Boot, Security Monitor [SM], Key Monitoring, Physical Unclonable Functions) but are still important for securing hardware accelerators. Developers and researchers pursue these different features to secure TEEs focused on hardware acceleration.  

\begin{figure}[t]
    \centering
    \includegraphics[width=1.0 \linewidth]{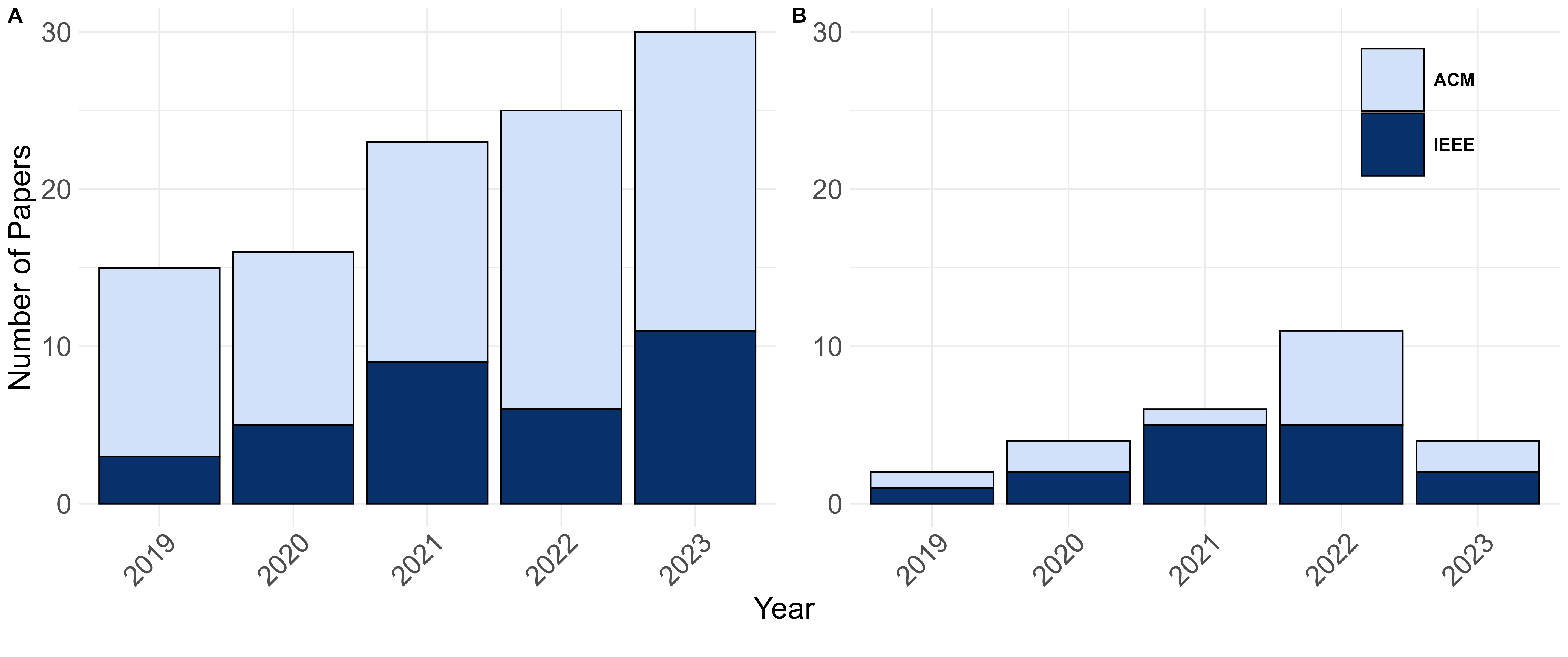}
    \caption{Stacked bar plot representing the number of papers published over the study period. Panel A is the 109 papers found using the search strings in Table \ref{table:strings}. Panel B is the 27 papers after application of the inclusion criteria in Table \ref{table:criteria}.}
    \label{fig:papers_over_time}
\end{figure}

\subsubsection{Cloud and Remote Computing}
\label{cloud&remote}
Papers on hardware acceleration almost always also focus on accelerators in a cloud computing environment (see references in Acceleration and Cloud Computing rows in Table \ref{table:applications&features}). Papers already discussed in Section \ref{accel} are re-mentioned but specific features are only discussed again here where relevant. Seven papers discuss custom-designed TEEs that implement security for cloud or remote-based FPGAs. Though applications are numerous for FPGA-based cloud computing, most papers found a need for security in a multi-tenant cloud environment. Key features such as attestation, memory security, enclaves, and RoTs are used to secure cloud environments that house accelerators \cite{cloud_issues}. 

Two papers discuss cloud and remote computing independent of hardware acceleration. Papers on \textit{MeetGo} \cite{MeetGo} and \textit{Operon} \cite{Operon} both provide TEEs for cloud and remote computing environments. \textit{MeetGo} is a hardware-centric solution to insider threats in cloud computing. \textit{MeetGo} implements a TEE that operates independently of the host systems architecture, restricting the administrator's access to users' data in the cloud. \textit{MeetGo's} modularity was demonstrated when it was implemented as a cryptocurrency wallet and General-Purpose Graphics Processing Unit \cite{MeetGo}. \textit{Operon} \cite{Operon} aims to provide secure, encrypted database operations while maintaining compatibility with existing SQL applications. Papers \cite{ShEF, AccGuard, AccShield, TrustToken, teleco} also are applied to cloud and remote computing, but have already been discussed in Section \ref{accel}.

\subsubsection{Attack Mitigation}
\label{attack}
Trusted Execution Environments play a critical role in attack mitigation. Almost one-fifth of the literature focuses on attack-specific mitigation through custom TEE implementation. Side channel attacks (SCAs) are a significant threat to TEEs. \textit{ChaosINTC} \cite{ChaosINTC} and \textit{REHAD} \cite{REHAD} both focus on SCA mitigation, interrupt-based and cache-based, respectively. \textit{ChaosINTC} implements a dynamic interrupt delay mechanism alongside an interrupt handler to protect their TEE \cite{ChaosINTC}. \textit{REHAD} uses reconfigurable hardware to mitigate cached SCAs \cite{REHAD}. While SCAs are a threat to TEEs specifically, TEEs are also used to defend against other threats.

The remaining TEEs discussed in the literature focused on preventing diverse attack vectors. \textit{TrustToken} features isolated execution and trusted user interaction to combat software-based assaults seeking information and unauthorized access \cite{TrustToken}. Yet another TEE seeks to combat unauthorized access, specifically through Trojans, by implementing a Hardware Trojan detection, identification, and recovery mechanism \cite{trojan}. Another attack vector, fault attacks, is mitigated by \textit{SecWalk}, which protects virtual and physical memory against fault attacks\cite{SecWalk}. From fault attacks to information leakage, TEEs often provide a first line of defense against bad actors. 

\subsubsection{IP Licensing}
\label{IP}
Of the papers that do not discuss hardware accelerators, cloud computing, and attack mitigation, there are a few niche applications. Intellectual Property (IP) protection and licensing is a concern for \cite{IP-Licensing} and \cite{TrustToken} because of multi-tenant environments. These multi-tenant FPGA environments present new security risks; current solutions necessitate third-party involvement for key-programming and encryption. The aforementioned \textit{TrustToken} \cite{TrustToken} only permits trustworthy connections between third-party IP and the rest of the SoC, while Khan et el. \cite{IP-Licensing} propose a Security framework for handling key storage and security monitoring. 

\subsubsection{Smart Grid Security}
\label{smart}
Smart Grid Security \cite{smart_grid} is a niche application that implements a TEE with dual-core isolation and secure boot based on a RoT. The niche applications of IP and grid security advance the field of SoC-FPGA-based TEEs, opening the door to apply TEEs to other computing areas. Applications of TEEs are slowly expanding as demonstrated by the papers centered around hardware accelerators, cloud computing, and attack mitigation. 

The application of TEEs across various domains, from hardware acceleration to cloud computing and attack mitigation, showcases their versatility and growing importance in securing modern computing environments. The innovative use of enclaves, attestation, and memory isolation in these environments highlights the challenges associated with maintaining security in dynamic, resource-shared settings. Meanwhile, the application of TEEs in attack mitigation, particularly against SCAs and hardware Trojans, underscores the necessity of security mechanisms that can preempt and neutralize threats. 

Although the focus on niche applications like IP licensing and smart grid security may seem specialized, these examples illustrate the broadening scope of TEE deployment. This trend reflects a growing recognition of the need for secure environments across all facets of computing, driving innovation and expansion in TEE capabilities.

\subsection{Feature Specific}
\label{features}

\begin{figure}[t]
    \centering
    \includegraphics[width=1.0 \linewidth]{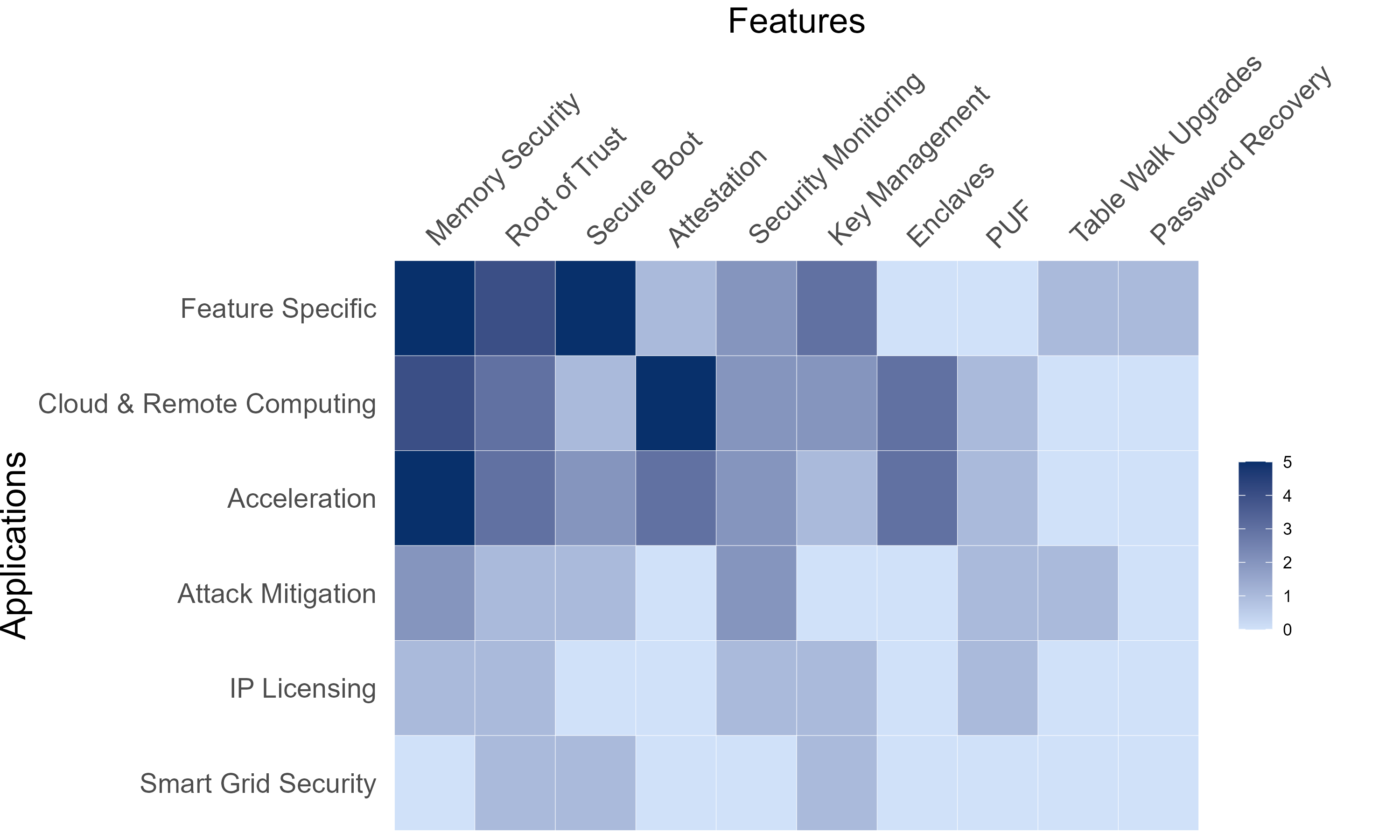}
    \caption{Heatmap of applications and their respective features in the pool of papers. Blue hue denotes number of papers discussing the features and applications indicated on axes.}
    \label{fig:heatmap}
\end{figure}

Nine papers focused on feature-specific TEEs. While all application-specific TEEs require a cadre of features, some researchers designed their TEEs with specific features in mind. These researchers put forth new contributions to the features TEEs can provide, however, not all features are implemented in tandem. These nine feature-specific papers focus on RoTs, attestation, memory security, secure boot, key management, and password recovery (Table \ref{table:applications&features}). Some papers focus on a singular feature, while others focus on multiple (Figure \ref{fig:heatmap}).

A RoT is a foundational element in most TEE designs, providing an anchor point for other security features, such as attestation and secure boot. Attestation ensures that the software and hardware components of a system are trustworthy. Mutual attestation allows devices of the same rank and type to verify their mutual interaction; Turan and Verbauwhede \cite{mutual_attestation} use a RoT to facilitate cryptographic verification, network communication, and decision-making, thus providing mutual attestation. Paper \cite{sec_hetero} employs a RoT, implemented using a Trusted Platform Module (TPM), as the basis for a protection-dedicated core in a multi-core RISC-V system. These feature-specific implementations show how RoTs provide the foundation for critical security features.

Memory security is crucial to the isolation of a TEE. Unrestricted or compromised memory access threatens entire system security. A notable memory-focused paper, \textit{ARES}, implements a security mechanism designed for non-volatile memory (NVM) in embedded systems \cite{ARES}. \cite{ARES} aims to combat common memory attacks and issues with Non-Volatile Memory (NVM) by implementing a novel Bonsai Merkle Tree (BMT) scheme and leveraging parallel recovery in FPGAs. Another NVM-focused TEE, \cite{NVM_prog}, proposes a methodology for securely booting from NVM in insecure environments, leveraging the reconfigurable logic of the FPGA as a secure anchor point. The Trusted Memory-Interface Unit sits in the reconfigurable logic region of the FPGA and performs integrity and authenticity verifications of NVM data prior to executing any user application, ensuring a secure boot process. The focus on NVM-based solutions highlights the importance of secure memory access in ensuring the integrity of data. 

\begin{table}[t]
    \centering
    \caption{Features and applications of Open-source, SoC-Based Trusted Execution Environments}
    \resizebox{\columnwidth}{!}{%
    \begin{tabular}{ll}
        \hline
        Topic & Paper \\
        \hline
        \textbf{Applications} & \\
        \hspace{4mm}Hardware Acceleration & \cite{ShEF, TACC, Software-defined, AccGuard, AccShield, TrustToken, teleco}\\
        \hspace{4mm}Attack Mitigation & \cite{TrustToken, ChaosINTC, SecWalk, REHAD, trojan}\\
        \hspace{4mm}Cloud and Remote Computing & \cite{ShEF, Operon, AccGuard, AccShield, TrustToken, teleco, MeetGo} \\
        \hspace{4mm}Feature Specific & \cite{mutual_attestation, key_scheduling, ARES, accel_page_walks, memory_encrpt, sec_hetero, Post-quantum, NVM_prog, password}\\
        \hspace{4mm}IP Licensing & \cite{IP-Licensing, TrustToken} \\
        \hspace{4mm}Smart Grid Security &  \cite{smart_grid}\\
        \hline
        \textbf{Features} & \\
        \hspace{4mm}Attestation & \cite{ShEF, Operon, mutual_attestation, AccGuard, AccShield, MeetGo} \\
        \hspace{4mm}Enclaves & \cite{ShEF, TACC, Operon, AccGuard} \\
        \hspace{4mm}Key Management & \cite{Operon, Post-quantum, IP-Licensing, AccGuard, key_scheduling, smart_grid, NVM_prog}\\
        \hspace{4mm}Memory Security & \cite{ShEF, TACC, ARES, Software-defined, accel_page_walks, SecWalk, AccShield, key_scheduling, TrustToken, NVM_prog, memory_encrpt, MeetGo}\\
        \hspace{4mm}Page Table Walk Upgrades & \cite{accel_page_walks, SecWalk}\\
        \hspace{4mm}Password Recovery & \cite{password}\\
        \hspace{4mm}Physical Unclonable Function & \cite{TrustToken}\\
        \hspace{4mm}Root of Trust & \cite{ShEF, Post-quantum, sec_hetero, mutual_attestation, AccGuard, key_scheduling, TrustToken, smart_grid}\\
        \hspace{4mm}Secure Boot & \cite{ShEF, TACC, sec_hetero, Post-quantum, mutual_attestation, key_scheduling, smart_grid, NVM_prog, ChaosINTC}\\
        \hspace{4mm}Security Monitoring & \cite{sec_hetero, IP-Licensing, AccGuard, TrustToken, trojan, password}\\
        \hline
    \end{tabular}%
    }
\label{table:applications&features}
\end{table}

Apart from NVM, memory encryption was the focus of a single paper \cite{memory_encrpt}. \cite{memory_encrpt} uses a special memory encryption unit that integrates directly with RISC-V architecture to encrypt memory using the lightweight ChaCha stream cipher which encrypts and decrypts quickly using the add-rotate-XOR (ARX) structure. Paper \cite{memory_encrpt} also utilizes the RISC-V Physical Memory Protection (PMP) unit to check load/store physical addresses against access restrictions. A spin-off of PMP presented by \cite{accel_page_walks}, Hybrid Physical Memory Protection (HPMP),  blends segment-based memory protection with a permission table, combining the strengths of both approaches. This hardware-software co-design dynamically manages memory protection and allocates segments and permission tables. These memory security approaches highlight the essential role of protecting memory in ensuring TEE security and integrity.

Secure boot is a critical feature in TEEs, ensuring that the system starts in a trusted state by verifying the authenticity and integrity of the bootloader and other essential components. The aforementioned \cite{NVM_prog} securely boots from NVM where the boot image is decrypted using the dynamically generated encryption key, and its integrity is verified by comparing the calculated hash against the stored token. Uniquely, \cite{Post-quantum} focuses on mitigating the threat of quantum computers on TEEs by implementing Secure Boot. The authors implement post-quantum secure boot using the eXtended Merkle Signature Scheme (XMSS) to protect the system's boot process from quantum computing attacks that could compromise traditional asymmetric cryptographic algorithms. This establishes a secure boot chain-of-trust from the RoT up to the operating system kernel ensuring the integrity of each boot stage \cite{Post-quantum}. 

In conjunction with secure boot, proper key management is essential to the security of TEE environments. Paper \cite{key_scheduling} proposes a novel approach to key management within the TEE by utilizing a flexible and secure boot procedure, complete isolation from the TEE domain, and exclusive secure storage for root keys. This ensures enhanced security and flexibility in key generation and maintenance. 

A few papers focus on less mainstream features such as password recovery and page table walk upgrades. \cite{password} implement a RISC-V processor, a secure coprocessor, and a password recovery engine connected through an AXI bus. The secure coprocessor includes an instruction set architecture (ISA) monitor and secure cache for secure computing tasks, especially those involving sensitive data like passwords \cite{password}. 

The diverse range of features explored across the literature highlights the components necessary for the deployment of TEEs in various computing contexts. The emphasis on foundational elements like RoTs and secure boot mechanisms underscores their role as the bedrock of secure system initialization and operation. These features establish and maintain trust, especially in environments where the integrity of both hardware and software must be assured. 

Memory security, with its various implementations, is particularly crucial given the pervasive risk of unauthorized access or data breaches that could compromise the entire TEE. However, the focus on specific features like password recovery and page table walk upgrades, though less common, reflects the growing complexity and specialization of TEE functionalities as they are adapted to meet the needs of increasingly diverse and demanding applications. This progression suggests that future research will push what TEES can achieve.

\subsection{Extensible TEEs}
\label{extensible}
The \textit{HECTOR-V} and \textit{Keystone} approaches provide modular and well-rounded TEEs, enabling users to plug and play rather than mix and match features and applications \cite{keystone, HECTOR-V}. 

\textit{HECTOR-V}, concerns itself with side-channel attacks, arguing that, ``TEEs, such as \textit{Intel SGX} or \textit{ARM TrustZone}, implemented on the main application processor, are insecure" \cite{HECTOR-V}. Focusing on combating SCAs, these authors implement a heterogeneous multicore architecture that embeds a dedicated processor into the system to separate the secure and non-secure domains. Their RISC-V Secure Co-Processor (RVSCP) restricts I/O access and provides control-flow integrity (CFI) for secure applications. This TEE provides secure I/O using identifier-based secure communication channels between different devices in the system, which ensures that only authorized entities can access sensitive peripherals. The RVSCP processor employs hardware-enforced CFI to safeguard applications running in \textit{HECTOR-V} using a specialized hardware unit to monitor the control flow of applications. Overall, \textit{HECTOR-V} aims to provide a secure architecture for trusted execution by combining a heterogeneous CPU architecture with secure coprocessor features, hardware control-flow integrity, and secure communication channels.

Lee et al. made a significant contribution to the TEE landscape when they created \textit{Keystone}, ``the first open-source framework for building customized TEEs" \cite{keystone}. \textit{Keystone} provides a comprehensive framework for implementing a modular TEE on an FPGA using RISC-V architecture. \textit{Keystone} TEEs use enclaves and (PMP) to isolate different computing modes from accessing data. While memory security is critical, it is not the only feature \textit{Keystone} TEEs provide. \textit{Keystone} TEEs also provide a configurable security monitor (SM) that adds a trusted layer below the OS that can be configured to enforce TEE guarantees (e.g., policies and security primitives). In addition to the SM, the secure boot and attestation capabilities measure and verify the integrity of the SM and enclaves. The myriad features are accompanied by SCA mitigation as \textit{Keystone} TEEs incorporate cache partitioning and other techniques to defend against side-channel attacks. In sum, Lee et al.'s comprehensive, open-source approach allows developers to have modularity and freedom when implementing and modifying a TEE created using the \textit{Keystone} framework. 

While both \cite{HECTOR-V} and \cite{keystone} present similar frameworks for TEEs, only one has been validated. The \textit{Keystone} framework, implemented by \cite{keystone-sensor}, served as the architecture for a trusted IoT sensing system. The sensing system features \textit{Keystone} and employs two types of Physically Unclonable Functions (PUFs)--one for the main device and one for the subordinate sensor. In this application, \textit{Keystone} provides isolation from potentially untrusted operating systems and applications using its enclave system. The \textit{Keystone} TEE integrates with a PUF, which serves as a hardware RoT that generates a unique, device-specific key for secure key management. This implementation of \textit{Keystone} illustrates how its modularity and feature-rich build allow multi-application realization. 

\textit{Keystone} and \textit{HECTOR-V} are easily adaptable to any chip using the RISC-V instruction set, though not without foibles. \textit{Keystone}, while highly modular and customizable, heavily relies on specific RISC-V hardware features i.e. PMP. Physical Memory Protection also limits the number of memory regions that can be protected based on PMP entries. \textit{HECTOR-V's} multicore architecture is complex to design and implement, particularly regarding the two communication between the cores. Along with the complex design, the hardware architecture and required resources of \textit{HECTOR-V} could limit its adaptability. Though these two TEEs use non-chip-specific features can be implemented across FPGA vendors, there are still some constraints when it comes to these frameworks.

Additionally, despite these advancements, several critical limitations persist that must be addressed. Performance overhead, particularly in memory encryption and secure boot processes, can slow down system operations, making TEEs less viable for resource-constrained environments like IoT devices and embedded systems, where efficiency is critical. Integration complexity, especially in heterogeneous architectures, complicates the seamless coordination between secure and non-secure domains, risking potential security gaps or performance bottlenecks. Additionally, while TEEs are designed to protect against many known threats, they remain vulnerable to emerging challenges such as quantum computing and advanced side-channel attacks. These limitations are crucial because they not only constrain the current utility of TEEs but also underscore the urgent need for ongoing research to develop more efficient, adaptable, and resilient security solutions.

\section{Threats to Validity}
We examine three potential threats to validity based on the classification scheme of \cite{cook1979quasi} and \cite{campbell1966experimental}.

Construct validity refers to how well the study identifies and categorizes TEEs. The search strings may have failed to capture relevant papers. This threat was mitigated by checking references of the included papers for potential oversights. Another threat is the manual categorization of papers (e.g., application-specific or feature-specific TEEs). This relies on subjective judgment. To mitigate this, possible features and applications were reviewed and rechecked.

Content validity may be affected in two ways. First, if the inclusion criteria used to select the final 27 papers were too restrictive, this would result in excluding papers that offer theoretical frameworks or nascent areas of research. This threat was minimized by reading the abstracts of all 109 papers to ensure no relevant studies were excluded. Second, only IEEE or ACM were searched, possibly excluding relevant papers published elsewhere. This is not a significant threat because IEEE and ACM conference proceedings and journals are the primary outlets for publications on edge-computing security.

External validity relates to the ability to generalize the findings of this study. We do not perceive significant threats to the external validity of this study. The scope of our study is on SoC-FPGA TEEs. Within this scope, our research captures the state of the published research. However, extrapolating or generalizing findings beyond this scope to the broader landscape of edge computing is not advised. 

\section{Conclusion}
This study systemizes SoC-FPGA TEEs, highlighting research gaps. Through the analysis of 109 papers sourced from IEEE Xplore and ACM Digital Library, a pool of 27 papers represented the current state of SoC-FPGA-based TEEs. These papers demonstrated the research challenges of implementing a robust, multi-featured, multi-application TEE, illustrated by the emphasis on application and feature-based TEEs. A robust, modular approach emerged in two papers combining critical features for a non-application-specific approach. The lack of publications related to SoC-FPGA-based TEEs that do not rely on third-party technology reveals a gap in the literature and an opportunity for researchers and developers (Figure \ref{fig:papers_over_time}). Many papers emphasize specific applications or features, but few combine features to create extensible TEEs. These insights hold significance for future development of TEEs and emphasize the importance of secure computing across applications and platforms. 

\section{Acknowledgments}
NASA and Resilient Computing, LLC supported this research under award number 80NSSC23CA147, and subcontract number 4W9082, respectively. This research was conducted with the U.S. Department of Homeland Security (DHS) Science and Technology Directorate (S\&T) under contract 70RSAT24KPM000022. Any opinions contained herein are those of the authors and do not necessarily reflect those of NASA, Resilient Computing, LLC or DHS S\&T. Thank you to Yvette Hastings at MSU for assisting with Figs. \ref{fig:papers_over_time} and \ref{fig:heatmap}.

\balance
\bibliographystyle{IEEEtran}
\bibliography{references}

\end{document}